\documentclass[reprint,aps,prx,superscriptaddress,amsmath,amssymb]{revtex4-1} 
\usepackage[T1]{fontenc}
\usepackage{color}
\usepackage{bm}
\usepackage{amsmath}
\usepackage{amssymb}
\usepackage{graphicx}
\usepackage{esint}
\usepackage{mathrsfs}
\usepackage{float}
\usepackage{wasysym}
\usepackage{amssymb}
\usepackage[title]{appendix}
\usepackage{natbib}
\usepackage{float}
\usepackage{tabularx}
\usepackage{diagbox}
\usepackage{fourier} 
\usepackage{array}
\usepackage{makecell}
\usepackage[table]{xcolor}
\usepackage{array}
\newcolumntype{P}[1]{>{\centering\arraybackslash}p{#1}}
\newcolumntype{M}[1]{>{\centering\arraybackslash}m{#1}}
\makeatletter

\makeatletter
\newcommand*{\rom}[1]{\expandafter\@slowromancap\romannumeral #1@}
\makeatother

\usepackage{times}{\Large }

\makeatother

\usepackage{babel}
\begin{document}
\title{Fast Quantum Calibration using Bayesian Optimization with State Parameter
Estimator for Non-Markovian Environment}

\author{Peng Qian}
\thanks{These authors contributed equally to this work.}
\affiliation{Beijing Academy of Quantum Information Sciences, Beijing 100193, China}

\author{Shahid Qamar}
\thanks{These authors contributed equally to this work.}
\affiliation{Beijing Academy of Quantum Information Sciences, Beijing 100193, China}

\author{Xiao Xiao}
\affiliation{Beijing Academy of Quantum Information Sciences, Beijing 100193, China}

\author{Yanwu Gu}
\affiliation{Beijing Academy of Quantum Information Sciences, Beijing 100193, China}

\author{Xudan Chai}
\affiliation{Beijing Academy of Quantum Information Sciences, Beijing 100193, China}

\author{Zhen Zhao}
\affiliation{State Key Laboratory of Low Dimensional Quantum Physics, Department of Physics, Tsinghua University, Beijing, 100084, China}

\author{Nicolo Forcellini}
\affiliation{Beijing Academy of Quantum Information Sciences, Beijing 100193, China}






\author{Dong E. Liu}
\email{Corresponding to: dongeliu@mail.tsinghua.edu.cn}
\affiliation{State Key Laboratory of Low Dimensional Quantum Physics, Department of Physics, Tsinghua University, Beijing, 100084, China}
\affiliation{Beijing Academy of Quantum Information Sciences, Beijing 100193, China}
\affiliation{Frontier Science Center for Quantum Information, Beijing 100184, China}

\begin{abstract}
As quantum systems expand in size and complexity, manual qubit characterization and gate optimization will be a non-scalable and time-consuming venture. Physical qubits have to be carefully calibrated because quantum processors are very sensitive to the external environment, with control hardware parameters slowly drifting during operation, affecting gate fidelity. Currently, existing calibration techniques require complex and lengthy measurements to independently control the different parameters of each gate and are unscalable to large quantum systems. Therefore, fully automated protocols with the desired functionalities are required to speed up the calibration process. This paper aims to propose single-qubit calibration of superconducting qubits under continuous weak measurements from a real physical experimental settings point of view. We propose a real-time optimal estimator of qubit states, which utilizes weak measurements and Bayesian optimization to find the optimal control pulses for gates’ design. Our numerical results demonstrate a significant reduction in the calibration process, obtaining a high gate fidelity. Using proposed estimator we estimated the qubit state with and without measurement noise and the estimation error between qubit state and the estimator state is less than $2 \%$. With this setup, we drive an approximated pi pulse with final fidelity of $99.28\%$. This shows that our proposed strategy is robust against the presence of measurement and environmental noise and can also be applicable for the calibration of many other quantum computation technologies. 
\end{abstract}

\maketitle

\section{Introduction}

With the achievement of the ultimate goal of fault-tolerance quantum computation, 
quantum computers will be able to solve many useful problems much more efficiently than classical supercomputers~\cite{Nielsen}. 
The main issue with the execution of quantum computational algorithms is that current quantum computers are very vulnerable to errors: 
a major source of errors is decoherence, which is due to the interactions between the encoded qubits and the external environment. 
While the effect of decoherence on quantum chips based on different hardware technology -superconductor qubits, semiconductor qubits, Rydberg atoms, etc.- can be more or less preeminent, it is a universal feature of current technology, which turn current quantum computers into Open Quantum Random Systems (OQRS) ~\cite{breuer2002theory, Gardiner}.
Therefore, decoherence needs to be understood and appropriately reckoned with.

In addition, for the correct functionality of quantum processors, all the qubits, quantum gates, and measurement controls need to be carefully and precisely calibrated \cite{majumder}. 
The calibration of a quantum processor includes state and gate fidelity estimation, the optimization of the system working parameters (such as qubit frequencies, coupler, and measurement resonator frequencies), as well as finding the optimal control parameters to construct high-fidelity quantum gates. 
In fact, inadequate calibrations, including inaccurate qubit optimizations and gate performance characterizations, are detrimental to the system and are sources of errors~\cite{Kelly, Barends, Corcoles, Rudinger, Rol, Patterson, White, Preskill}. 

Decoherence effects, caused by the environment, can be categorized as Markovian~\cite{breuer2002theory} and non-Markovian processes~\cite{DeVega2017dynamics, breuer2016colloquium}. In a Markovian process, which is memory-less, decoherence dynamics are such that information exchanged between the system and the environment at a certain point in time does not influence future processes. On the other hand, for non-Markovian processes, due to memory effects, information can be repeatedly transferred between the quantum system and the environment, providing feedback effects and influencing the next stages of the dynamics. Hence, in non-Markovian processes, the dynamics are generally more complex \cite{QiBo, Yamamoto}.
A challenging but unavoidable task is to find a quantum calibration protocol that can take non-Markovian environmental influences into account.


Recently, major breakthroughs have been achieved in superconducting quantum processors, where the attainment of quantum ``supremacy” (or ``advantage”) has been claimed~\cite{Gong-USTC2021-Science,  Arute, WuYulin, boixo2018,aaronson2016,bouland2019,Pino-TrapIon2021-Science}. 
For state-of-the-art superconducting processors with extremely short gate times of less than 100 ns~\cite{Barends}, the qubit lifetime/gate time ratio reaches a value close to 1000~\cite{BarendsRami2014}.  
These advances imply that we entered the so-called noisy intermediate-scale quantum (NISQ) era of quantum computers~\cite{Preskill}. 
However, high-fidelity operations of NISQ and larger quantum computers can only be accomplished through the arduous task of engineering accurate quantum processor calibration protocols. 
Optimal calibration relies on the accurate understanding of every component and estimation of each parameter in the quantum processor, including working frequencies of the resonators and qubits, qubit coupling coefficients, pulse control parameters, along with the qubit state or fidelity estimations~\cite{DiCarlo, Yamamoto2010, Mariantoni, Yang}.

At the heart of quantum calibration lie optimization algorithms, designed to maximize the fidelity of the system over the input field for both quantum state preparation and gate realization. 
The standard techniques to detect and minimize systematic errors, such as quantum fidelity estimation~\cite{eisert}, demand perfect measurements and state preparation along with high measurement and sample complexities~\cite{chow2009}.   
\begin{figure*}[t]
\includegraphics[width=2\columnwidth]{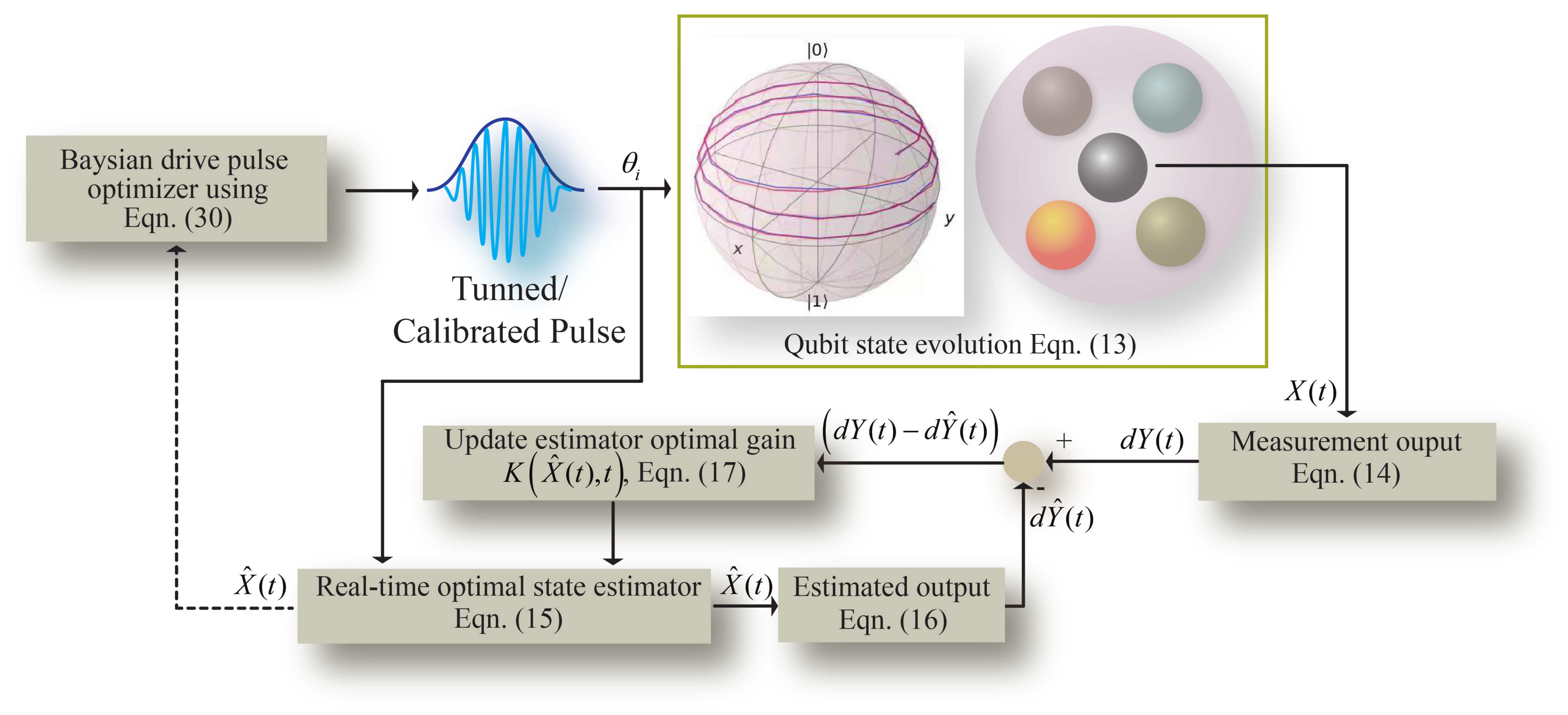}
\caption{Fast qubit optimal calibration process for state estimation and gate design of a superconducting qubit: To monitor the dynamics $X(t)$ of the real system, the probing fields $u_x(t)$ and $u_y(t)$ are injected into the qubit system to move $X(t)$ to the desired value, the output of which is measured by $Y(t)$. $dY(t)$ is used to drive ROSE to estimate the state $\hat{X}(t)$ of the real system using optimal gain matrix $K(\hat{X}(t),t)$, to minimize the distance between the estimator and the system. The control scenario $f(\theta)$ is predicted and iterated based on available estimated data. Its performance needs to be evaluated experimentally. The optimization task is to find the optimal control parameters $\theta_i$. The Bayesian optimizer recommends the next set of parameters for the qubit system based on new update evaluations at each step.}\label{fig:cal_diag}
\end{figure*}
Many methods have been suggested and implemented to achieve efficient quantum state estimation, including randomized benchmarking~\cite{chen2016,magesan}, linear regression ~\cite{qi2013}, compressed sensing ~\cite{rodionov,Kalev} and maximum likelihood ~\cite{blume}. 
However, the feasibility of these schemes breaks down when the system dimension becomes large.
Considering the time overhead of the calibration sequence, repeating the sequence from scratch is far from ideal. Thus, complex calibrations and parameter drifts can be achieved, in the case of small systems, with precise manual tuning. 
However, it is clear that any form of calibration based on manual control is not scalable, and therefore a fully autonomous solution is needed for the operation of systems with a large number of qubits. 

From the perspective of quantum measurement applied to calibration, the main difficulty is that the relevant parameters cannot be identified by a single measurement; however, continuous monitoring of the quantum systems allows for coherence to be maintained ~\cite{Kim}. 
Unlike classical systems, quantum states are affected by dephasing and spontaneous decay, even when taking continuous weak measurements, and are also affected by backaction of the measurements. Such effects need to be taken into account ~\cite{braginsky,wiseman,Jacobs2006}. 
However, by continuously monitoring the qubits, the Hamiltonian of the system can be modified in real-time to obtain some desired outcome ~\cite{Hacohen,foroozani}. 
The use of continuous monitoring systems also opens up new avenues for the estimation and control of quantum systems, such as state estimation ~\cite{Nielsen,Vijay,Bisio,Ladd}, parameter estimation using Bayesian optimization ~\cite{Yang,kiilerich,Bejanin,Korotkov} and compressed sensing methods ~\cite{rodionov,Kalev}.
Because of these advantages, continuous measurement protocols are being applied in quantum computing~\cite{Bisio,Ladd,zhang2017}, quantum information~\cite{qin2017}, 
parameter estimation~\cite{kiilerich}, quantum measurement techniques, and superconducting circuits applications~\cite{Vijay,kjaergaard}. 


Optimization algorithms are playing a vital role in the quantum system calibration ~\cite{riaz,Shahid,gerster2021,sivak2022}, and are broadly categorized by their ability to calculate the gradient of the parameters requiring optimization. 
Gradient-based algorithms~\cite{krotov,Caneva} and hybrid quantum-classical methods~\cite{Li,Sauvage,Egger} can be readily applied in experiments. 
However, these methods suffer from the presence of local minima ~\cite{Zahedinejad2014,Bukov,Sauvage}.
Alternatively, gradient-free algorithms such as Nelder-Mead~\cite{Poggiali,Kelly2014,Rach, Crow}, evolutionary algorithms~\cite{Zahedinejad2014, costa2021, Pantita} and reinforcement learning techniques \cite{sivak2022,niu} are more global than gradient-based methods, but they require many iterations to find optimal solutions. 
In this regard, Bayesian Optimization (BO)~\cite{martinez,brochu,shahriari} is a non-gradient based method that offers an attractive alternative, as it can efficiently choose a new set of parameters at each step,  and it is applicable to quantum optimization~\cite{zhu,nakamura, mukherjee,Sauvage}. 
BO can significantly reduce the number of iterations needed to achieve convergence, and therefore can be used to further speed-up global optimization. 
Therefore, it is particularly suitable for fast calibration in quantum experiments, since it requires fewer evaluations to optimize the drive pulse and reduces the number of runs, all of which makes the calibration process much faster.


\begin{center}
\textbf{{Summary of results}}
\end{center}

In this work, we propose a fast qubit gate calibration scheme based on a newly developed state parameter estimation method for non-Markovian environments, along with Bayes’ optimization for gate control pulse optimization. The method is schematically shown in Fig.~\ref{fig:cal_diag}. 
Quantum state estimation and control pulse optimization are crucial for the full characterization of quantum states since the required resources grow exponentially with the number of qubits. Our method estimates the state in real-time, and the purpose of the optimal real-time algorithm is to detect the value of the real system using continuous measurement data. 
In this context, the term ``real-time" means that measurements made during the implementation of the algorithm inform subsequent measurements, which reduces the total number of measurements needed.
Subsequently, the real-time estimator refines the estimates of state parameters at each measurement step and can be successfully used to increase the information obtained with each measurement, thereby reducing the required calibration resources and time. 

In order to check whether our technique can be implemented in real qubit systems, we performed an experiment with superconducting qubits. 
We can see that with our real-time optimal state estimation, which we name ROSE, together with BO, we can significantly shorten the calibration time of superconducting qubits. 
We believe that our method will improve present medium-scale superconducting quantum computers and will also scale up to larger systems. 
Our results are mainly applicable to the tune-up step in quantum information processing, i.e., the initial calibration and maintenance of the system.
Our algorithm employs a modern Bayesian inference strategy to opt for informative experimental settings for gate design, while maintaining computational feasibility, and numerical results show robustness to experimental imperfections. 
The real-time fast qubit optimal calibration process for state estimation and gate design of a superconducting qubit is shown in Fig. \ref{fig:cal_diag}.

Notably, this proposed Bayesian optimization based on our ROSE approach is not limited to superconducting qubits but can also be useful for other multi-qubit systems and quantum error correction codes, as it only depends on applications of non-Markovian single-qubit calibration operations and measurements.
Hence, the results of this paper are suitable for optimizing other systematic errors occurring in other physical quantum processing platforms, such as Rydberg atoms~\cite{Crow,Saffman}, semiconductor qubits \cite{zhang2019,takeda}, trapped ions \cite{Lange,blatt,nigg,nautrup} or in solid-state structures~\cite{hanson,gambetta,fedorov}.

The paper is organized as follows: in section II we describe the single-qubit model. Section III describes ROSE as applied to superconducting qubit states. 
Section IV discusses how Bayes optimization fits into the general framework of superconducting qubit calibration, with regards to the number of evaluations required to optimize the drive pulse for a given qubit gate design. In section V our numerical results are discussed in detail. 
Section VI concludes the paper with a brief outlook on future work.

\section{System Description}

In general, the calibration of single superconducting qubit is divided into the following steps: 1) Detect and verify the basic information of the qubit; 2) find the initial attempt point of the control parameters of the quantum gate according to these information; 3) apply these parameters to electronic equipment to generate a gate to certain known initial state; 4) readout and estimate the final state to obtain the fidelity of the gate; 5) according to the fidelity estimation, adjust the control parameters of the gate to find improved parameters; finally we need to repeat step 3) to step 5) to optimize the control parameters until the fidelity of the gate reaches a certain expectation.

\begin{figure}[H]
\centering
\includegraphics[width=1\columnwidth]{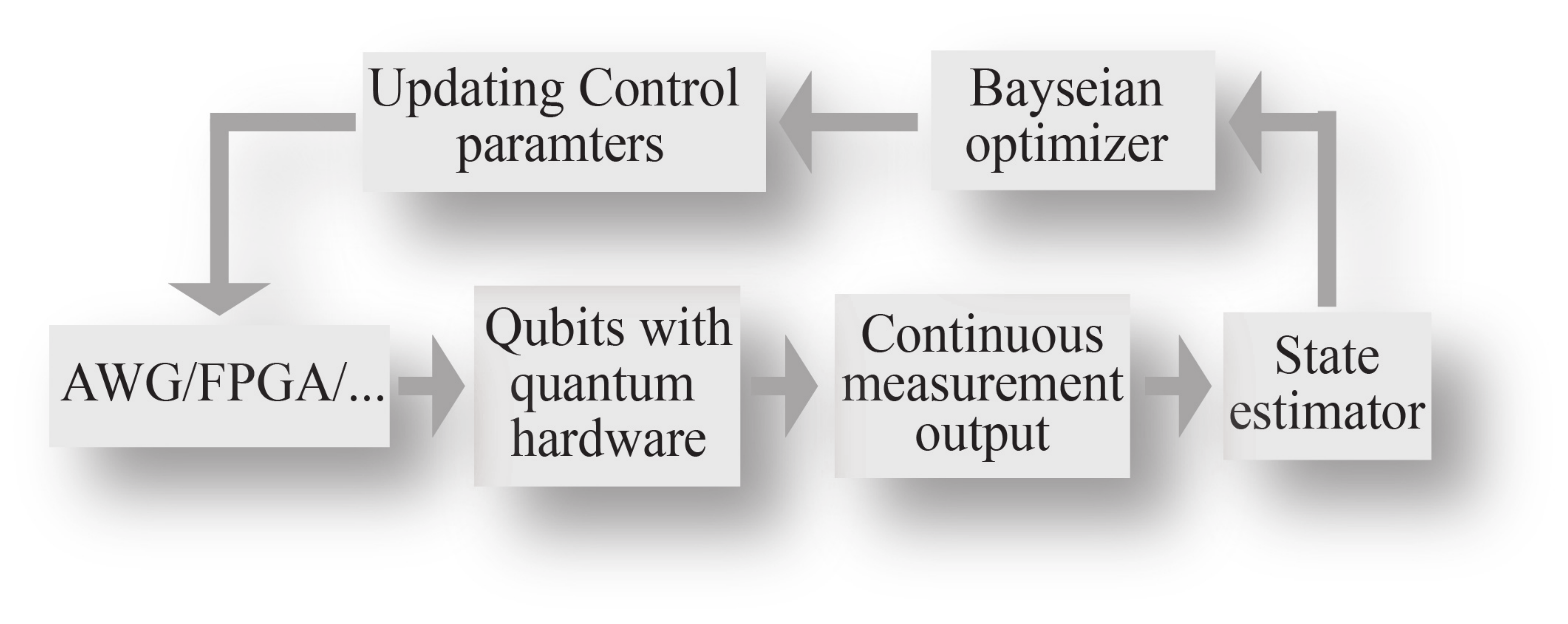}
\caption{Superconducting qubit calibration process}\label{Cali_Procs}
\end{figure}

In order to extract the information of the gate fidelity for the calibration procedure, the one-shot measurement need to be repeated for many times in, for an example, standard randomized benchmarking~\cite{chen2016,magesan}. However, this procedure is time-consuming and becomes less efficient especially for the large-scale qubit system calibration task, which needs to call a large number of single qubit calibration procedures. Therefore, a novel fast and accurate fidelity estimation scheme is necessary for the calibration task of quantum processors. Based on this motivation, we will introduce an optimal method using continuous weak measurement to speed up the calibration process to achieve the high fidelity gate. 
Before discussing the optimal estimator and pulse optimizer in the later sections, we first discuss the theoretical framework for our scheme--the noisy quantum control along with continuous weak measurement. The process of superconducting qubit is depicted in Fig.~\ref{Cali_Procs}.

The dynamics of single qubit system which interacts with the environment can be described by a stochastic quantum master equation~\cite{Nielsen,zhang2017,altafini}
\begin{equation}\label{eqn:4}
\begin{aligned}
d\rho(t)=&-i[H_s,\rho(t)]dt+(-i[H_c(t),\rho(t)])dt\\&+L_1(t)[\rho(t)]dt+L_2(t)[\rho(t)]
\end{aligned}
\end{equation}
where $\rho(t)$ is the density matrix of the qubit system, the single qubit Hamiltonian is $H_s=$ $\omega_0 \sigma_z$, $\omega_0$ is the energy of system, $H_c(t)=\sum_{k=1}^2u_k(t)\sigma_k=u_x(t)\sigma_x+u_y(t)\sigma_y$ describes the control part, where $\sigma_x$, $\sigma_y$ and $\sigma_z$ are the Pauli matrices. $L_1 [\rho(t)]dt$ depicts the Lindblad term  and $L_2 [\rho(t)]$ shows the stochastic term.
The Lindblad term $L_1 [\rho(t) ]dt$ can be described as
\begin{equation}\label{L1}
\begin{aligned}
L_1(t)[\rho(t)]dt=&[\Delta (t)+\gamma(t)]\mathcal{D}[\sigma^-]\rho (t) dt \\ &+[\Delta t(t)-\gamma(t)]\mathcal{D}[\sigma^+]\rho (t) dt \\&+M\mathcal{D}[\sigma_z]\rho(t) dt.
\end{aligned}
\end{equation}
where $\sigma^-=(\sigma_x-i\sigma_y)$ and $\sigma^+=(\sigma_x+i\sigma_y)$ are the lowering and raising operators. The parameters $\Delta(t)$ and $\gamma(t)$ represent the diffusion and dissipation coefficients of the quantum system.  They come from time integral of an environment spectrum $J(\omega)$~\cite{maniscalco,Cui}:
 \begin{equation}
J(\omega)\propto\omega\frac{\omega^{2}_{c}}{\omega^{2}_{c}+\omega^{2}}
\end{equation}
which is an ohmic spectral density with a Lorentz-Drude cutoff function, and $\omega_c$ is the high cutoff frequency.
 \begin{equation}
\gamma(t)=2\int_{0}^{t}d\tau\cos(\omega_{0}\tau)\int_{0}^{\infty}d\tau'J(\omega)\coth\left(\frac{\hbar\omega}{2k_{B}T}\right)\cos(\omega\tau')
\end{equation}
 \begin{equation}
\Delta(t)=\int_{0}^{t}d\tau\sin(\omega_{0}\tau)\int_{0}^{\infty}d\tau'J(\omega)\sin(\omega\tau')
\end{equation}
where $\omega$ is the bath frequency.
 Then the analytical expression $\gamma(t)$  can be expressed as~\cite{maniscalco}
\begin{equation}
\gamma \left( t \right)=\frac{{{\alpha_c }^{2}_c}{{\omega }_{0}}{{r}^{2}}}{1+{{r}^{2}}}\left[ 1-{{e}^{-r{{\omega }_{0}}t}}\cos \left( {{\omega }_{0}}t \right)-r{{{e}^{-r{{\omega }_{0}}t}}}sin\left( {{\omega }_{0}}t \right) \right],
\label{eq:gamma_disp}
\end{equation}
where $\alpha_c$ is the coupling constant, and $r$ is the ratio of $\omega_c$ and $\omega_0$. The diffusion coefficient $\Delta(t)$ has the form,
\begin{equation}
\Delta \left( t \right)=2{{\alpha_c }^{2}_c}{{k}_{B}}T\frac{{{r}^{2}}}{1+{{r}^{2}}}\left\{ 1-{{e}^{-r{{\omega }_{0}}t}}\left[ \cos \left( {{\omega }_{0}}t \right)-\frac{1}{r}\text{sin}\left( {{\omega }_{0}}t \right) \right] \right\}
\end{equation}
where $k_B T$ depicts environmental temperature. The parameters $\Delta(t)$ and $\gamma(t)$  contain very important Markovian and non-Markovian  features of open quantum systems. The indispensable difference between Markovian systems and non-Markovian systems is the presence of environmental memory effect. Due to system reservoir interactions the system loses quantum information into the environment. Apparently, if the environment has a non-trivial structure then the lost information can go back to the system in future time, which leads to the non-Markovian dynamics with memory~\cite{Cui}.

It is important to note that the dissipative term comprises of two decoherence channels in a non-Markovian open quantum system. The first two terms in Eq. (\ref{L1}) illustrate the energy from the environment. The last term labels the phase-shift process, which has no energy process for energy exchange with the environment and thus has no impact on the population of ground and excited states, but mitigates the off-diagonal elements in the density matrix \cite{Roch}. 

The $L_2 [\rho(t)]$ term has the form
\begin{equation}\label{eqn:6}
{{L}_{2}}\left[ {{\rho }{(t)}} \right]=\sqrt{M\eta }\mathcal{H}\left[ {{\sigma }_{z}} \right]{{\rho }{(t)}}d{{W}{(t)}},
\end{equation}
where $0<\eta<1$ is the detection efficiency and $M\geq0$ is the interaction strength between the system. $W(t)$ is the Weiner process noise which satisfies the It\^{o}'s rules, $E[dW(t)]=0$, $[dW(t)]^2=dt$ and can be described
\begin{equation}\label{eqn9}
dY(t)=tr\left( {{\sigma }_{z}}{{\rho }{(t)}} \right)dt+\frac{dW(t)}{\sqrt{M\eta}}
\end{equation}
where $dY_t$ is the output of the observation process.

Then dynamics of system (\ref{eqn:4}) can be written as

\begin{eqnarray}\label{eqn:7}
\begin{aligned}
d{{\rho }{(t)}} =&-i{{\omega }_{0}}\left[ {{\sigma }_{z}},~{{\rho }{(t)}} \right]dt-i{{u}_{x}}\left( t \right)\left[ {{\sigma }_{x}},~{{\rho }{(t)}} \right]dt\\&-i{{u}_{y}}\left( t \right)\left[ {{\sigma }_{y}},~{{\rho }{(t)}} \right]dt+\left[ \Delta \left( t \right)+\gamma \left( t \right) \right]\mathcal{D}\left[ {{\sigma }^{-}} \right]{{\rho }{(t)}}dt
\\ &+\left[ \Delta \left( t \right)-\gamma \left( t \right) \right]~\mathcal{D}\left[ {{\sigma }^{+}} \right]{{\rho }{(t)}}dt+M\mathcal{D}\left[ {{\sigma }_{z}} \right]{{\rho }{(t)}}dt~\\&+\sqrt{M\eta }\mathcal{H}\left[ {{\sigma }_{z}} \right]{{\rho }{(t)}}d{{W}{(t)}},
\end{aligned}
\end{eqnarray}
 The value of measurement strength $M$ is below the  control strength which is the amplitude of the pulse $|u_x|$, e.g. $0.1$ MHz to $10$ MHz for a typical range.

The superoperators $\mathcal{D}$ and $\mathcal{H}$ are
\begin{eqnarray}\nonumber
\mathcal{D}\left[ {{\sigma }_{z}} \right]{{\rho }{(t)}}={{\sigma }_{z}}{{\rho }{(t)}}\sigma _{z}^{\dagger }-\frac{1}{2}\sigma _{z}^{\dagger }{{\sigma }_{z}}{{\rho }{(t)}}-\frac{1}{2}{{\rho }{(t)}}\sigma _{z}^{\dagger }{{\sigma }_{z}},\\
\mathcal{H}\left[ {{\sigma }_{z}} \right]{{\rho }{(t)}}={{\sigma }_{z}}{{\rho }{(t)}}+{{\rho }{(t)}}{{\sigma }_{z}}-\left[ tr\left( {{\sigma }_{z}}{{\rho }{(t)}}+{{\rho }{(t)}}{{\sigma }_{z}} \right) \right]{{\rho }{(t)}}\nonumber.
\end{eqnarray}
The density matrix $\rho(t)$ of a two-level quantum system can be
defined by $[x(t),y(t),z(t)]$ in the Cartesian coordinate system
as $\rho(t)=\frac{1}{2}(I+x\sigma_{x}+y\sigma_{y}+z\sigma_{z}).$ The
stochastic master equation (SME) of non-markovian qubit with continous
measurement in Bloch sphere representation has the form
\begin{equation}\label{eqn:14}
\begin{aligned}
dx(t)=&-\left(\Delta(t)+\frac{M}{2}\right)x(t)dt-\omega_{0}y(t)+u_{y}(t)z(t)dt\\ &-\sqrt{M\eta}x(t)z(t)dW(t)\\
dy(t)=&-\left(\Delta(t)+\frac{M}{2}\right)y(t)dt+\omega_{0}x(t)-u_{x}(t)z(t)dt\\ &-\sqrt{M\eta}y(t)z(t)dW(t)\\
dz(t)=&-2\gamma(t)dt-2\Delta(t)z(t)dt-u_{y}(t)x(t)dt\\ & +u_{x}(t)y(t)dt-\sqrt{M\eta}(z^{2}(t)-1)dW(t)
\end{aligned}
\end{equation}
 
We use $\omega_0=E$ to represent energy of the system. The quantum Weiner noise after interaction with the system is the
output of the quantum system. The stochastic differential equation
of the output of quantum system can also be described as
\begin{equation}
dY(t)=z(t)dt+\frac{dW(t)}{\sqrt{M\eta}}\label{output}
\end{equation}
where $Y(t)$ is the measurement process output of continuous weak
measurement.

From equation (\ref{eqn9}), one can see that the measurement data is not only affected by measurement noise but also contains some information about the state of the system. By measuring the output of the system, the information about the quantum state can be extracted. Therefore, the output of the system can be processed to design state estimators and control pulse optimization for the gate design of qubit. Consider the following dynamical system equivalent to the SME and output measurement process,
\begin{equation}\label{qubitdy}
dX(t)=A_{0}(t)dt+A_{2}(t)X(t)dt+G(X(t))dW(t)
\end{equation}
\begin{equation}\label{output}
dY(t)=CX(t)+\frac{dW(t)}{\sqrt{M\eta}}
\end{equation}
where $X(t)=[x(t),y(t),z(t)]^{T}\in\{X(t)\in R^{3}||X(t)|\le1\}$
are vector form of Bloch sphere. $A_{0}(t),A_{2}(t),G(X(t)),C$ are
matrix:
\begin{equation}\nonumber
\begin{aligned}
&A_{0}(t)=\left[\begin{array}{c}
0\\
0\\
-2\gamma(t)
\end{array}\right],~~C=[0,0,1],\\
&A_{2}(t)=\left[\begin{array}{ccc}
-\Delta(t)-\frac{M}{2} & -\omega_{0} & u_{y}(t)\\
\omega_{0} & -\Delta(t)-\frac{M}{2} & -u_{x}(t)\\
-u_{y}(t) & u_{x}(t) & -2\Delta(t)
\end{array}\right],\\
&G(X(t))=\left[\begin{array}{c}
-\sqrt{M\eta}x(t)z(t)\\
-\sqrt{M\eta}y(t)z(t)\\
-\sqrt{M\eta}(z^{2}(t)-1)
\end{array}\right]
\end{aligned}
\end{equation}
With these equations, we have a full description of qubit dynamics in a non-Markovian environment with continuous weak measurement. 


As we know that using measurement output data we can get only a part of information about the qubit state. In that sense, to extract the complete fidelity information, we will propose an optimal estimator for the qubit state in order to monitor the dynamics of the real system described by the Eq. (\ref{qubitdy}) at each particular trajectory. The measurement signal $dY(t)$ will serve as the input of the optimal estimator to estimate the complete dynamic state variables of the real system (\ref{qubitdy}). Further, using this estimator we estimate trajectory of qubit's pure state, i.e., trajectory having control and environmental affect but without considering measurement affect.

\section{Real-time Optimal State Estimator}

One important step in quantum calibration is to estimate the state of qubit from the measurement output. Many estimation schemes for quantum states are designed recently~\cite{Wieczorek,Xue,amini,vuglar}. However, those linear estimators are designed for the linear quantum system models, i.e., $X(t)=A_2(t)X(t)+W(t)$, where $A_2(t)$ and $W(t)$ is the state matrix and noise as shown in Eq. (\ref{qubitdy}). In this case, one can see that the state $X(t)$ and the noise $W(t)$ are linearly independent. Those linear state estimators only work for such type of linear quantum system dynamics. But when there is nonlinearities in the system's state, for example, the matrix $G(X(t))$ in Eq. (\ref{qubitdy}) contains the information about qubit state and is coupled/multiplicative with the noise $W(t)$ as shown in our system Eq. (\ref{qubitdy}), which results in a nonlinear state equation. Therefore, the state of such system cannot be estimated using linear state estimators.


Thus, a nonlinear real-time estimator needs to be developed to estimate the observable values for the system in Eq. (\ref{qubitdy}) using the measurement output signal from the principal quantum system. 
The optimal estimator based on the state-dependent differential Riccati equation (SDDRE) is one of the highly promising real-time estimation strategies for the such nonlinear systems~\cite{Dani,Shahid}. The real-time estimator can be designed in the presence of state-dependent multiplicative noise with a state-dependent coefficient form~\cite{Cimen,Qamar}. The estimator is connected to the output of the real system, whose dynamics can be determined by minimizing the mean square deviation between the system and the estimator state in real-time to achieve the desired calibration outcomes. 

In this section, we propose a nonlinear Real-time Optimal State Estimator (ROSE) which is designed to reduce the measurement complexity and improve the qubit calibration efficiency. The ROSE includes the estimated behavior of the quantum system for reconstructing the dynamical state variables of the qubit system using measurement data. In this scheme, the weak measurement provides an alternate method to estimate quantum states. During the measurement, the target information can be extracted without significantly disturbing its state by using a continuous weak measurement (CWM).
However, the state $\rho(t)$ cannot be fully estimated with a single measurement record $Y(t)$, because the operator associated with $Y(t))$ holds partial information about the state of the system. Hence, this is an optimal estimation problem, and for this, the output $Y(t)$ is used to drive the state estimator to estimate the system state. It is assumed that all system parameters except the state are known and that the state $\rho(t)$ can be accurately estimated by the ROSE using SDDRE. The qubit ROSE is designed by using SDDRE and estimator gain is dependent on $\hat{X}(t)$.

By considering the system Eq. (\ref{qubitdy}) and measurement output Eq. (\ref{output}), here the goal is to design the ROSE to estimate $\hat{X}(t)$ of the system state $X(t)$ that decreases the mean square error, i.e., $E\left[\|X(t)-\hat{X}(t)\|\right]^2$ at time $t$  from a sequence of noisy measurement data $\left[Y(s):t_{0}\le s\le t\right]$. This gives $\hat{X}(t)=E\left[{X}(t)\vert\lbrace Y(s):t_{0}\le s\le t\rbrace\right]$, i.e., the conditional expectation value of $X(t)$ given complete measurement data $Y(t)$. We construct the estimator by adding the correction term $K_{1}(\hat{X}(t),t)(dY(t)-d\hat{Y}(t))$ into the estimator's dynamical equation with the same structure of the system dynamics described by Eq. (\ref{qubitdy}) to ensure the realization of state estimation. Here, $K_{1}(\hat{X}(t),t)$ is the ROSE gain matrix, $dY(t)-d\hat{Y}(t)$ is the innovated error between the system output and that of ROSE, which describes the information gain from output $Y(t)$, $dY(t)$ is the actual measurement output of the system, and $d\hat{Y}(t)$ is the estimated output of the ROSE. Thus, the proposed ROSE in this paper has the form
\begin{equation}\label{estimated system}
\begin{aligned}
d\hat{X}(t)=&A_{0}(t)dt+A_{2}(t)\hat{X}(t)dt\\&+K_{1}(\hat{X}(t),t)(dY(t)-d\hat{Y}(t))
\end{aligned}
\end{equation}
where
\begin{equation}\label{estimated output}
d\hat{Y}(t)=C\hat{X}(t)dt.
\end{equation}
Here, we define an estimator of system's state $\hat{X}(t)=[\hat{x}(t),\hat{y}(t),\hat{z}(t)]$. The design purpose of the ROSE is to find the estimator gain $K_{1}(\hat{X}(t),t)$ that minimizes the estimation error covariance matrix $P_{1}(\hat{X}(t),t)$, where $P_{1}(\hat{X}(t),t)=E[e_{1}(t)e_{1}(t)^{T}]$, $e_1(t)=X(t)-\hat{X}(t)$ and E{[}\ensuremath{\centerdot}{]} is the expectation function. If the optimal gain matrix $K_{1}(\hat{X}(t),t)$ is designed properly, the error between estimated state $\hat{X}(t)$ and real state ${X}(t)$ will be close to zero. Satisfying the above requirements means that the system is asymptotically stable. After minimizing the error covariance, i.e., $\min\limits_{K_1(\hat{X}(t), t)} \ tr \ \dot{P}_{(\hat{X}(t), t)}$ w.r.t., $K_1(\hat{X}(t), t)$ (see Appendix A),  we reach the optimal gain matrix $K_1(\hat{X}(t), t)$ as:
\begin{equation}
K_{1}(\hat{X}(t), t)C)=P_{1}(\hat{X}(t), t)C^{T}M+G(\hat{X}(t))\sqrt{M}
\end{equation}
where ${P}_{1}(\hat{X}(t),t)$ is the solution of state dependent differential Riccati (SDDRE) equation and has the form
\begin{equation}\label{CovarianP}
\begin{aligned}
\dot{P}_{1}(\hat{X}(t), t)=&(A_{2}(t)-G(\hat{X}(t))C\sqrt{M})P_{1}(\hat{X}(t), t)\\&+P_{1}(A_{2}(t)-G(\hat{X}(t), t)C\sqrt{M})^{T}\\&-P_{1}(\hat{X}(t), t)C^{T}CP_{1}(\hat{X}(t), t)M
\end{aligned}
\end{equation}

It can be seen from Eq. (\ref{CovarianP}) that SDDRE is a function of state variables $\hat{X}(t)$, which can be computed by integrating with initial conditions of $P_{1_0}(\hat{X}(t), t )$ of above Eqn. (\ref{CovarianP}). Due to the state-dependent from of ${P}_{1}(\hat{X}(t)$, the calculation must be done online (here "online" means that measurements are taken during the execution of ROSE to update the later ones) at each update step to continuously obtain a real-time estimation of the system. As, the measurement output $dY(t)$ appears in the estimator Eqn.~(\ref{estimated system}) indicates that the estimator drives based on the actual output measured at each time step. Therefore, ROSE (\ref{estimated system}) is useful in a real-time experimental setting, because, it updates the state in real-time. Based on the ROSE we propose a Bayesian optimization method to optimize the drive pulse to achieve a high fidelity gate in a very short-time, and to enhance the calibration fidelity and efficiency for a superconducting qubit.



\section{Gate Design based on Bayesian Optimization}
\begin{figure}[t]
\centering
\includegraphics[width=1\columnwidth]{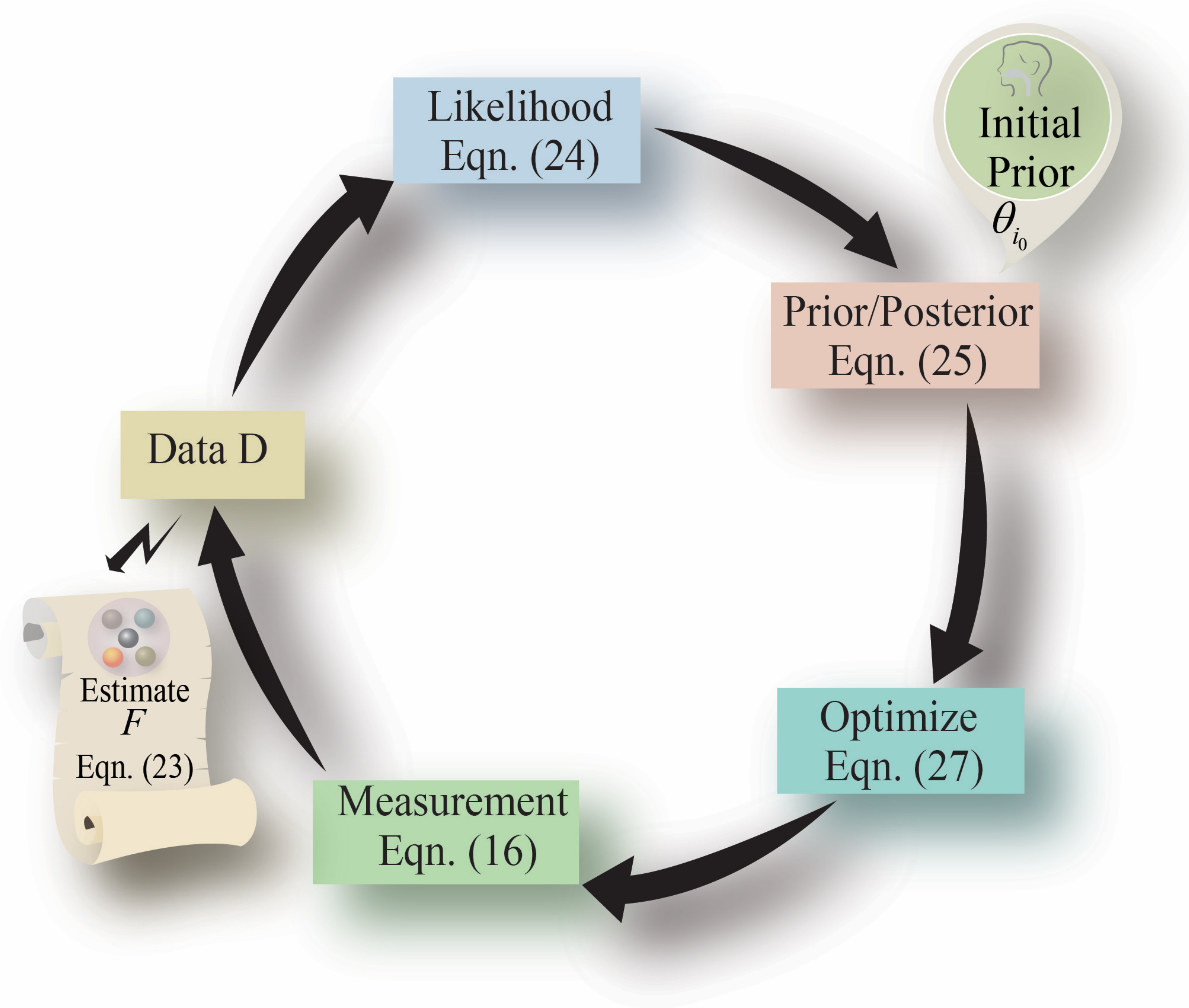}
\caption{Bayesian optimization algorithm}\label{Bayes_opt}
\end{figure}
The core of any quantum optimal control framework is an optimization algorithm, which aims to maximize the "\textit{Figure of Merit}" (FoM) of an experiment regarding the input control field. Here, FoM measures the performance or efficiency of an experiment when accomplishing a particular task, e.g. achieving a desired quantum gate. In typical quantum experiments, the input control field has some adjustable parameters, such as the amplitude and phase of magnetic/electric fields, and parameters to control the microwave pulses. There are methods~\cite{Kelly2014, Rach,riaz,rodionov} proposed to find fast optimization protocols. Their performance is limited regarding the number of the data points collected from the control landscape for the estimation task, because, they need large parameter’s search space and large numbers of iterations with the requirement of large data set.

There is no doubt that one can improve the optimal solution by examining more points on the control landscape for the optimization algorithm. But, if FoM is derived from a quantum experiment, the complexity of the experiment procedures and their time complexity becomes exponentially large as the system size increases. This limitation also applies to the numerical calculation tasks in any calibration protocols where the size of the Hilbert space increases exponentially with the system. In addition, due to experimental noise including preparation and measurement noise, the uncertainty and errors cannot be exactly removed when evaluating FoM, and this fact causes the optimization process more challenging.

In these scenarios, the Bayesian optimization (BO) is a smart and attractive tool to find out the optimal parameters of the input control field to get the high fidelity gate of the qubit at a fast pace based on limited data set with less number of interactions required. The BO is based on a gradient-free approach that provides an choice because at each step of the optimization process it chooses the next set of parameters to be evaluated. This can result in a significant reduction in the number of iterations required for convergence when performing global optimization. BO has the additional advantage of integrating probabilistic elements of data collection, which can be used to further improve its efficiency even in the presence of noise~\cite{Sauvage,brochu,mukherjee}.
Therefore, we propose a Bayesian optimization to regulate pulse parameters to design a superconducting qubit gate. 
The schematic diagram of Bayesian optimizer is depicted in Fig. \ref{Bayes_opt}.

For the superconducting quantum computing, the qubit gates are controlled by microwave pulses which has the form
\begin{equation}
    u(t)=\cos(w_d t)\epsilon_x(t)+\sin(w_d t)\epsilon_y(t), \ \ t<t_g
\end{equation}
where $t_g$ represents the length of time of control, $w_d$ denotes the frequency of microwaves, and, $u_x (t)=\cos(w_{d} t)\epsilon_x(t)$ and $u_y (t)=\sin(w_{d} t)\epsilon_y(t)$ is the x and y components from an arbitrary wave generator. 
The x-components are usually taken into a Gaussian waveform, i.e.,
\begin{equation}
    \epsilon_x(t)=A\exp\bigg[-\frac{(t-t_g/2)^2}{\sigma^2}\bigg]+B
\end{equation}
Through the analysis of Derivative removal via adiabatic gate (DRAG) scheme~\cite{motzoi2009}, one can set the y-component waveform as
\begin{equation}\label{contr1}
\epsilon_y(t)=\frac{\alpha_{s}}{\delta}\cdot\frac{d\epsilon_x(t)}{dt}
\end{equation}
where $\delta$ is the non-harmonicity of the superconducting qubit, which is defined as the energy difference $\Delta\epsilon_{2-1}-\Delta\epsilon_{1-0}$ where $\Delta\epsilon_{i-j}$ is the energy difference between the $i^{th}$ and the $j^{th}$ energy levels.
Here we need to optimize the parameter $\alpha_{s}$ in $u(t)$ to estimate the optimal value of control pulse. The purpose of DRAG is to counter or reduce the qubit leakage processes generated by being excited to a higher energy level. In addition to the leakage errors, the phase errors can produce detrimental effects in this scheme. To reach a trade-off between two types of errors, one can introduce additional control parameter $\phi$ via~\cite{chen2016}:
\begin{equation}\label{contr2}
u(t)\rightarrow {u}_{1}(t)=u(t)e^{i2\phi t}.
\end{equation}
In this scheme, we have two control parameters $\alpha_{s}$ and $\phi$, which need to be optimized for achieving a high-fidelity gate, where the gate fidelity is defined as
\begin{equation}\label{gate fi}
F=\frac{1}{6}\sum_{j=\pm x,\pm y, \pm z} \mathrm{Tr}[U_{\text{ideal}}\rho_j U^\dagger_{\text{ideal}}M_p(\rho_j)]
\end{equation}
Here, $U_{\text{ideal}}$ is the ideal unitary gate, and $M_p(\rho_j)$ is a quantum  channel describing the  actual under-calibrated and noisy gate.
We consider the average gate fidelity on the different initial states by averaging over case with the six eigenstates of the three different Pauli operators. The state obtained by the gate is due to manipulation of these eigenstates and is a manifestation of the fidelity of the gate operation.

We use FoM (short for "indicator", abbreviated as $f$) to describe how well a goal is achieved. $f$'s dependency on control parameters defines the control scenario. In our case, $f$ is just the gate fidelity of Eq.~(\ref{gate fi}). In the spirit of Bayesian analysis, we can preset and construct many different control scenarios and assign a probability to each preset control (a different dependence on parameters). 
Control scenario $f$ is predicted and iterated based on available information. Its performance needs to be evaluated experimentally. The general literature refers to these $f$ as proxy models~\cite{Sauvage,shahriari,mukherjee}. This represents some of our possible estimations and substitutions for the real-world scenarios.

As the FoM can be acquired numerically or experimentally and passed to the optimizer. 
During the calibration procedure, the FoM is acquired from computational evaluations using experimental data, and then the information is input into the optimizer for further processing. These evaluations can be time-consuming and subject to noise in experimental settings. To perfectly resolve the true optimization landscape on a suitable probabilistic model, $f$ is specified in terms of distribution function $P(f)$, i.e., Gaussian process (see Appendix B). $P(f)$ is referred to as prior distribution, which allows the 
use of regular and smooth functions to illustrate the control landscape.

Furthermore, BO updates $p(f)$ based on the already collected values of the FoM. At an arbitrary step $D$ of the optimization, the vector $D$ is obtain by set of measurement data and is denoted by a vector $[f(\theta_0),...,f(\theta_i),...,f(\theta_m)]$. The posterior distribution $P[f(\theta|\theta_i,D)]$ defined as the distribution for $f$ conditioned on all the data $D$ obtained so far.
Moreover, $P(D)$ is the probability distribution of observation set $D$ and $P(D|f)$ is the likelihood of observation set $D$ with given $f$ (see Appendix B). i.e.,
\begin{equation}\label{likehood}
    \text{log}(P(D|V,I))=-\frac{1}{2}D^Tk^{-1}D-\frac{1}{2}\text{log}|k|-c
\end{equation}
where, $V$ and $I$ are the mean and variance and, hyper-parameters of the kernel function $k$. We determine $P(f)$, so that we can predict the $f$-value of $f(\theta)$ for control parameters $\theta\rightarrow{(\alpha_{s},\phi)}$ that has not yet been measured. At the same time, we can theoretically deduce it
\begin{equation}
    P(f(\theta|\theta_i,D))=\mathcal{N}(\bar{f},\textrm{cov}(f))
\end{equation}
where $\mathcal{N}$ depicts the Gaussian distribution and
\begin{eqnarray}\label{upd12}
\begin{aligned}
\bar{f}=&k(\theta,\theta_i)k^{-1}(\theta_i,\theta_j)D\\
\textrm{cov}(f)=&k(\theta,\theta)-k(\theta,\theta_i)k^{-1}(\theta_i,\theta_j)k(\theta_i,\theta)
\end{aligned}
\end{eqnarray}
where $\bar{f}$ represents the average of $f(\theta)$ under the new parameters $\theta$ and $\textrm{cov}(f)$ is the variance.

\begin{figure*}[!t]
\includegraphics[width=2\columnwidth] {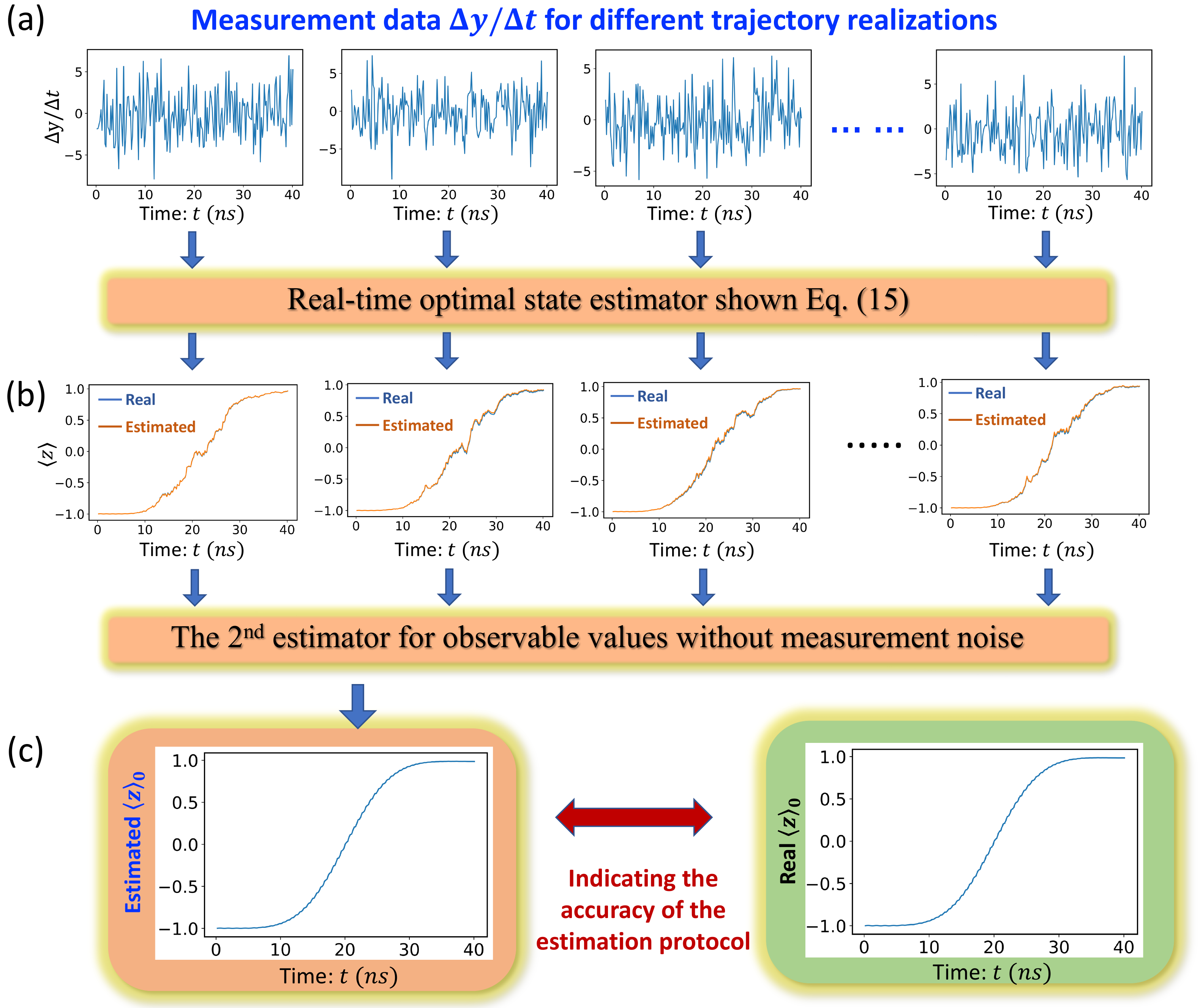}
\caption{Data transformation to the estimated state by estimator. (a) By taking several examples of measurement data $\frac{\Delta y}{\Delta t}$, in which $\Delta t$ is the time interval between measurement points, and these data contain deterministic part and random part due to the uncertainty principle. $ns$ is the nanosecond for the time unit. (b) According to these data, we apply our first part of the estimator to track the estimated trajectory of the qubit and compared it with the real one.
(c) We apply the second part of our estimator to infer the observable values $\langle z\rangle_0$ of the qubit without measurement noise}\label{cal1}
\end{figure*}

Finally, it is also necessary to determine which set of control parameters $\theta$ to use in the next step of iterative optimization. Thus, we compute an acquisition function $\textrm{ac}(\theta)$ and next set of parameters $\theta$ is chosen when $\textrm{ac}(\theta)$ reaches at its maximum value. The acquisition function $\textrm{ac}(\theta)$ has the form
\begin{eqnarray}\label{acfun1}
\textrm{ac}(\theta) = \beta \ast \textrm{cov}(f)+f(\theta)
\end{eqnarray}
where $\beta$ determines the convergence speed and the local area size. The larger the $\beta$, the less likely it is to fall into local extremums, and the smaller the $\beta$, the faster the convergence. We can dynamically adjust the $\beta$, take the larger value during the initial search, and then slowly begin to decline to a smaller value.

Numerically, we look for the argument $\theta$ that makes the acquisition function $\textrm{ac}(\theta)$ larger. This process does not require experimental operations, only numerical optimization is needed. We take the search field segments, i.e., using the maximum method in each segment to find the largest $\textrm{ac}(\theta)$, and then we take a maximum of these $\textrm{ac}(\theta)$. This $\theta$ is the point we need to determine for our next experiment. We measured that $f(\theta)$ to obtain new data points, so that the data set become $[f(\theta_0),...,f(\theta_i),...f(\theta_m)]$. At the same time, we can predict new $f(\theta)$ and calculate the new $\bar{f}$ and $\textrm{cov}(f)$ according to Eq. \eqref{upd12}. The new acquisition function $\textrm{ac}(\theta) =\beta\ast \textrm{cov}(f)+f(\theta)$ is then calculated. Here the parameter $\beta$ will slowly decrease to around $0$ as this iteration progresses.

For gate calibration, we can take $f\rightarrow{F}$ and ours interest focus on optimization of $\theta\rightarrow{(\alpha_{s},\phi)}$. We initially choose randomly several groups of $((\alpha_{s})_i,\phi_{i})$ and apply the corresponding gate and use the estimator to get $F((\alpha_{s})_i,\phi_{i})$. Then we transport these data to Bayesian optimizer and have a suitable model $f$ and numerically get the optimal of acquisition function $\textrm{ac}(\alpha_{s},\phi)$ and point to next try parameter $(\alpha_{s},\phi)$. Then we apply new $(\alpha_{s},\phi)$ iteratively to generate a gate until we achieve high fidelity.

\section{Numerical results and discussion}

\begin{figure}
\centering
\includegraphics[width=\columnwidth]{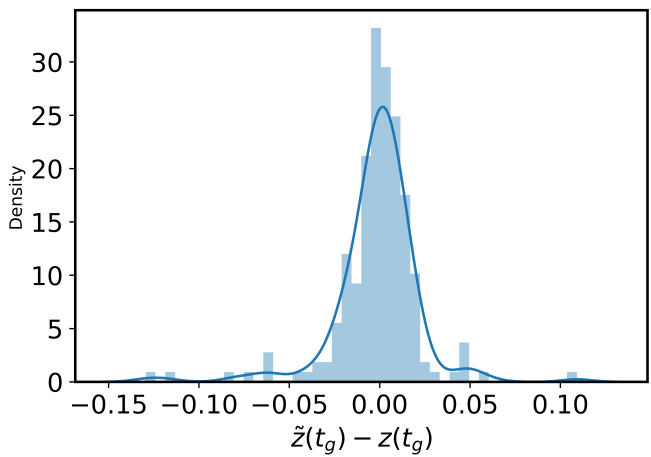}
\caption{Final state fidelity error counting for 200 trajectories. We define $z(t_{g})$ as the real state fidelity and $\Tilde{z}(t_{g})$ as the estimated one for end of each trajectory. We show the distribution of this fidelity error for 200 different trajectories}\label{distrib}
\end{figure}

This section gives the numerical simulation results for the comparison between the real system simulation and proposed estimation strategy given in above sections II-IV(A), and want to show that our strategy is valid  and efficient. We apply Bayesian optimizer based on ROSE for the gate calibration of superconducting single qubit system with non-Markovian environments. Here, we assume that the qubit dynamic simulation based on the Eq.~(\ref{qubitdy}) corresponds to the real physical system. It generates the output as shown in Eq. (\ref{output}). This is the only information obtained in the
real physical experiments. While we can estimate the information of simulating system (real system) using ROSE, i.e. Eq.~(\ref{estimated system}), those information can not be directly obtained from  physical experiments. In addition, we assume that the real experiments obey the dynamics  Eq.~(\ref{eqn:14}) and also with $M=0$. Additionally, during continuous measurements, the amount of information extracted from the measurement data approaches to zero as the duration of the measurement time tends to zero. To construct such measurements, we divide the time into several intervals, and the length of each interval $\Delta t$; and then consider a weak measurement process for every interval.



We now show our numerical results to indicate how well the proposed estimator ROSE tracks the true value of the state of the system. In our estimation and experiment/simulation, we assume that we know the qubit energy $\omega_0=E$ in estimator and measurement strength $M$, which means the system energy input for the estimator is the same as the real system. This can be realized by a other independent physical calibration process like Ramsey or spectrum scan. The flow of this section is as follows. We start from Fig. \ref{cal1}(a) which corresponds to the outcome $\Delta Y/\Delta t$ from the continuous measurement (averaged over a time interval $\Delta t$), and this measurement outcomes are generated by simulating the real experimental systems. The outcome contains partial information about the qubit state; and the information is covered by a stronger noisy signal at the small-time interval regime, because the information part is linear in time $z(t)dt$ while $dW(t)\sim \sqrt{dt}$. This outcome signal will be input to two estimators:1).  \textit{read-time optimal estimator} and 2).\textit{Second estimator for observable values without measurement noise}, to obtain Fig. \ref{cal1}(b) and Fig. \ref{cal1}(c) respectively. Then, we will also consider the cases: 3). \textit{State estimation with different energy parameter}, and 4). \textit{Effect of correlation time for non-Markovian environment}. Finally, we will consider calibration performance for the gate design protocol:5). \textit{Estimation of the control pulse parameters} and 6). \textit{Comparison with DRAG protocol}.


\textit{1). Real-time optimal estimator}: In Fig.~\ref{cal1}(b), we show some comparisons of the estimated state and real state. The blue line stands for the evolution of a real state while the orange one stands for our estimation. Here we take measurement strength $M=2$ MHz and two parameters of the environment are $\alpha_c=0.5$ and $r=0.01$. Furthermore, in this setup, we drive an approximated $\pi$ pulse using Baysian optimizer in Eq.~(\ref{acfun1}) with the gate fidelity of $99.28 \%$. We see in this setup, some trajectories of the state are close to the final state while some are not because of a relatively small $M$. Also, the estimated and real states are almost the same in each trajectory, and the error is less than $2 \%$. We show the distribution of errors in one sample in Fig.~\ref{distrib}. We see that the final state fidelity error between the estimated one and the real one in most trajectories is in the range of $2\%$. These results show the robustness of our method.

\textit{2). Second estimator without measurement-induced noise}: Using trajectories of ROSE in Fig.~\ref{cal1}(b), we implement the second estimator termed as "Pure state estimator (PSE)", which obtains the observable values without measurement noise. Since the measurement is only the tool for estimating the gate fidelity in the calibration task; therefore, the effect from the measurement-induced noise (measurement noise for short) should be eliminated in order to extract the essential information of the fidelity. The structure of this PSE is similar to ROSE.
To design the state estimator without measurement noise, first we take derivative of $\hat{X}(t)$ w.r.t., measurement strength $M$, i.e., $\frac{\partial \hat{X}(t)}{\partial M}=X(t)_p$ and estimate it by using Eqn.~(\ref{second_X}) [for more details, see Appendix B], 
where $\hat{X}(t)$ is the result of Fig.~\ref{cal1}(b) and $\hat{X}(t)-MX_p(t)=X(t)_0$ is shown in Fig.~\ref{cal1}(c). To estimate the state without measurement noise of system in Eq.~(\ref{qubitdy}) both estimator (ROSE and PSE) must be executed simultaneously, because, the state of PSE depends on ROSE [see Appendix B]. The Fig.~\ref{cal1}(c) we average of $200$ trajectories, and we get an average estimated pure state (PSE) with the fidelity of $99.26 \%$ compared to the real pure state, i.e., $99.28 \%$. The error between real pure and PSE state is less than $0.02 \%$.

\textit{3). State estimation with different energy parameter $E$}: 
Now, we will discuss the parameter range in which our proposed method works. Previously, we assumed that we have well knowledge about the system's energy $E$, like in the traditional calibration process (the Ramsey experiment). However, we cannot guarantee that each time we have an accurate $E$ of the system. Here we test with a different range of $E$ to see the efficiency of our method even though we do not know the value of the real system's energy $E$ exactly. We tune the energy parameter in the estimator and keep the energy parameter fixed in the real system to see the real and the estimated trajectories of PSE.
\begin{figure}
\centering
\includegraphics[width=1.1\columnwidth]{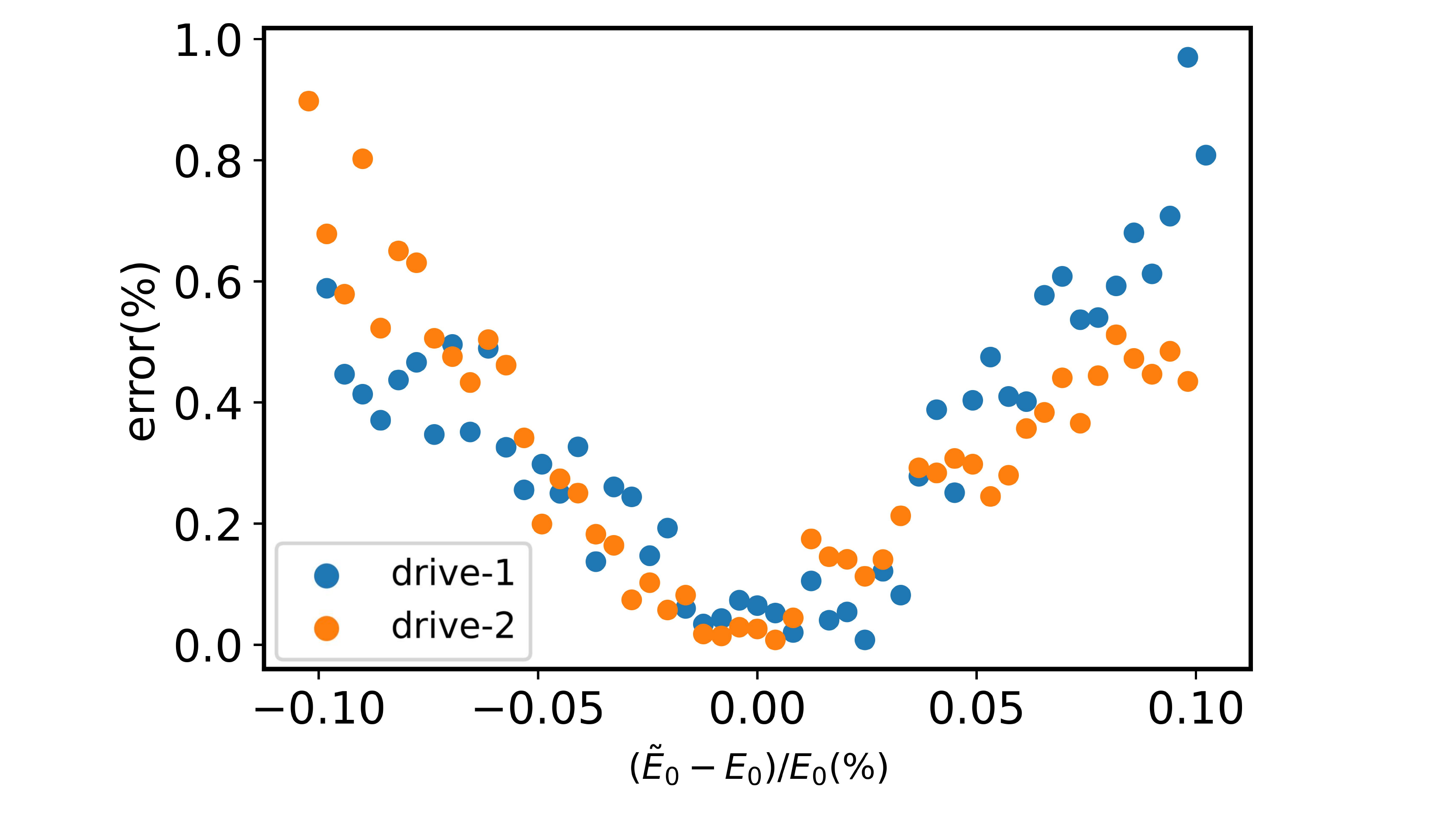}
\caption{Energy parameter induced error. This error indicates the ratio of the fidelity difference (between the estimation value and the real value) to the the real fidelity. We consider the estimation error as changing the misknowledge of the real qubit energy $E$. $E_{0}$ represents real one. $\Tilde{E}_{0}$ represents energy set in the estimator. The label "drive-1" and "drive-2" represent two different pulse drives with the different drive frequency.}\label{Diff_E}
\end{figure}

In Fig.~\ref{Diff_E}, the horizontal axis labels the difference between the energy parameter in the estimator and the simulating one (real one). We take the real energy value of the qubit as $4.88$ GHz and tune the qubit energy parameter in the estimator from $4.88$ GHz to $4.85$ GHz and each point step of $0.1$ MHz. We see near the range of $3$ Mhz difference, the error between the real state and our estimation is less than $1\%$. If we have a large error of more than $10$ Mhz in the system's energy, then the prediction of our method is a bit less accurate. In this result, we take measurement strength as $M=1.5$ MHz and the environment parameters are as $\alpha_{c}=0.25$ and $r=0.01$.
\begin{figure}
\centering
\includegraphics[width=\columnwidth]{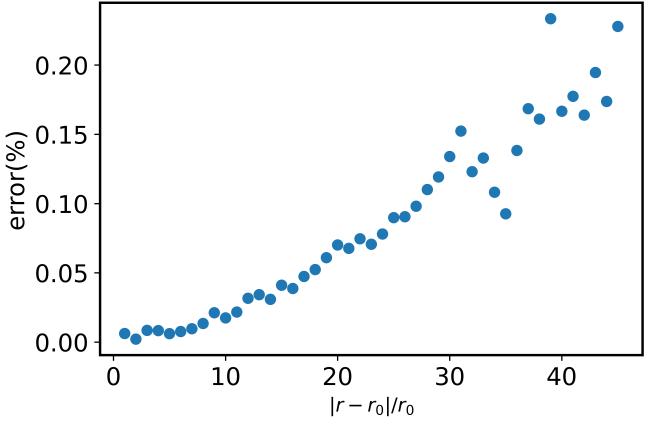}
\caption{Environment parameter induced error. The "error" here is defined the same way as in Fig.\ref{Diff_E}, but here We consider the estimation error as changing the misknowledge of the environment parameter. $r$ represents real one and we fix $r_{0}$ for the estimator where $r$ is defined in Eq.~(\ref{eq:gamma_disp}).}\label{Environt}
\end{figure}

\textit{4). Effect of correlation time for non-Markovian environment}: Additionally, we consider the effect of the correlation time of the non-markovian environment. When the time correlation becomes large, the non-markovian effect will become important and our model will suffer from some errors from the missing knowledge about the environment. Here we fix environment parameter in estimator as $\alpha_{c}=0.5$ and $r_{0}=0.01$. Then we change the environment parameter for real system(simulator) $r$ (defined in Eq.~\ref{eq:gamma_disp}) from this point $r_{0}$ to a smaller one until $\frac{r_{0}}{2}$ and thus we have a longer correlation and have a strong non-markovian effect for a real system. We see the difference between our estimated final pure state with the real final pure state. We see error increases as system parameters differ more from the estimator one. At most, it may lead to an error of $0.2 \%$. In Fig.~\ref{Environt}, we see that with the correlation time increasing, the error becomes large. Although in the case of small coupling between environment and system this error can be ignored, for a relative amount of coupling, we need to know the environment well first.

\begin{figure}
\includegraphics[width=\columnwidth] {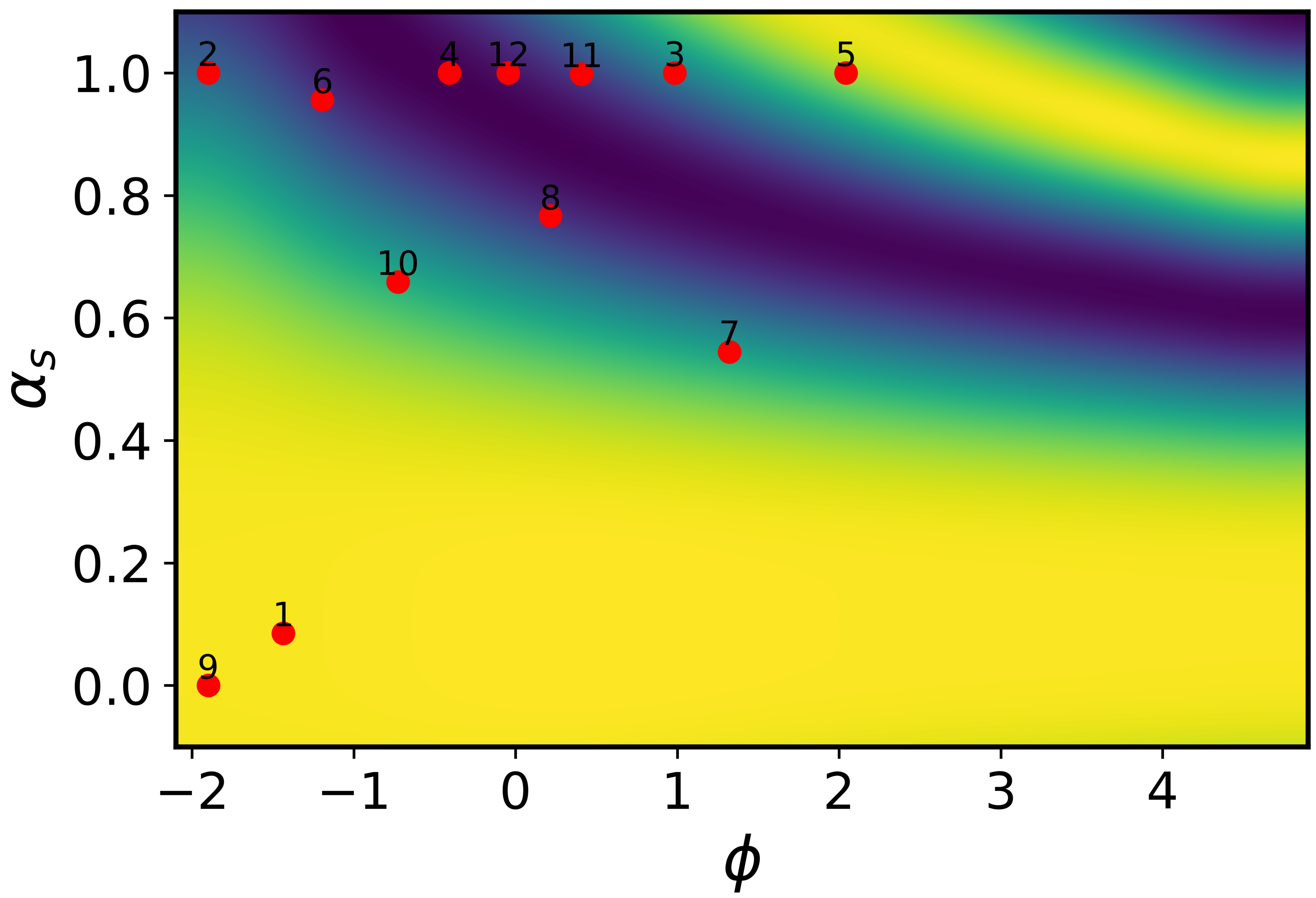}
\caption{Bayesian optimization for two parameters $\alpha_{s}$ and $\phi$. Here, we show how acquisition function $\textrm{ac}(\theta)$ optimizes $\alpha_s$ and $\phi$. The numbers in the red dots indicate the sequence of $\textrm{ac}(\theta)$. The initial point is labeled as "1" and the final one is labeled as "12" . The deeper color means higher fidelity.}\label{bayes}
\end{figure}

\textit{5. Control pulse parameters $\alpha_{s}$ and $\phi$}: Bayesian optimization algorithm works remarkably well to get the high fidelity gate in less number of iterations. Implementation of the Bayesian scheme is easier and computationally economical, flexible, and has fast convergence.  Finally, we talk about how the Bayesian optimization method works. In previous steps, we have got the accuracy of fidelity of the final state for the corresponding control pulse through our estimator. Now we proceed to find the optimal value of the control pulse parameter, i.e., $\alpha_{s}$ and $\phi$ in equations (\ref{contr1}) and (\ref{contr2}). First, we randomly choose several pairs of $(\alpha_{s}, \phi)$  to simulate/ run different experiments and generate data for each of them. Then we estimate the fidelity $F$ for the final state with our estimator for corresponding control parameters $\alpha_{s}$ and $\phi, F$. We take these control parameters and corresponding estimated fidelity $\alpha_{s}, \phi$ as an input of the Bayesian optimizer. Then the optimizer constructs a suitable model and predicts a pair of $\alpha'_{s},\phi'$ as optimized control parameters according to equation (\ref{acfun1}). We apply this predicted $\alpha'_{s},\phi'$ to simulate the system to generate data and estimate fidelity $F'$ for this control parameter pair. And we add $\alpha'_{s},\phi',F'$ to the beginning of the random sample $\alpha_{s},\phi,F$ and update Bayesian model. Then it will predict a new optimized control parameter. And we loop the previous steps until we get the desired fidelity. In Fig. \ref{bayes}, we show fidelity with different parameters by blue surface and use red points to show the predicted ones by the Bayesian optimizer. In this example, we use $10$ random initial tries. In these experiments, the drive frequency is $4.90$ GHz and the qubit frequency is $4.889$ GHz. We see that through the $15$ prediction we almost get the optimized point for fidelity.

\textit{6. Comparison with DRAG}
As we have optimized pair of control parameters in practice using Bayesian optimizer. In the test, the number of times we searched is generally in the order of 10 (for the accuracy of the control parameters of the experiment). In general, these two parameters are experimentally $0.01$ with MHz precision, while the search domain is roughly [-1, 0] and [-100, 0] MHz and the total parametric search point is the order of 1e5. The DRAG scheme~\cite{motzoi2009} attempts to minimize gating time through better control have so far had limited success. Because the DRAG strategy search separately and it roughly requires the squared number of times in a single parameter search. While Bayesian optimization is a method that infers the result by limited data and has a quick path due to its optimize structure. We only need relatively few data points. It is observed that the Bayesian efficiency is improved in both experimental resources and time as shown in table~\ref{tabel1}. Table~\ref{tabel1} depicts that the proposed Bayesian optimization based on a real-time estimator outer-performed than the traditional thousands of time repetitive measurement to probe the fidelity of qubit and drive pulse optimization using the traditional DRAG method. The DRAG scheme required large number of data points and time to search the pulse parameters than our proposed strategy. This means that our proposed method not only reduces the number of measurement data required to speed up the calibration process but also requires less time to achieve high fidelity compared to the traditional methods.

\begin{table}[h]
    \centering
    \caption{Comparison of gate calibration of Bayesian based on state estimation with DRAG pulse optimization and traditional repeative measurement to probe fidelity }
\begin{tabular}{|l|c|c|p||}\hline
\rowcolor{green!20} \backslashbox{Comparison\\ terms}{Methods}
&\vtop{\hbox{\strut Repetitive measurement +}\hbox{\strut DRAG pulse tuning}}&\vtop{\hbox{\strut Proposed} \hbox{\strut method}}\\\hline\hline
 Data number required &$10^5$ &$10^2$\\\hline
Test time for fidelity & $10^4 \mu$s & $10 \mu$s\\\hline
\vtop{\hbox{\strut Trying times for finding}\hbox{\strut optimal control}} &Search all & Tens try\\\hline
\end{tabular}\label{tabel1}
\end{table}

\section{Conclusion and Future Work}
To summarize, we proposed a straightforward and viable automated protocol for qubit gate calibration using Bayesian optimization, which is based on the real-time estimation of qubit states in superconducting circuits, allowing for the correction of systematic errors. Traditional calibration techniques are time-inefficient since they require tens of thousands of data points to minimize estimation error.

Our Bayesian optimization method adopts  what we call the
real-time optimal state estimation (ROSE) strategy for qubit gates. It is robust and efficient against noise measurements,
and it estimates specific errors efficiently and accurately using minimal resources, such as number of measurements data compared to traditional methods. This means, under typical conditions, an order of magnitude reduction in run-time: it makes use of real-time calibration, exploiting the error estimates obtained in one experiment to reduce the errors in the subsequent experiment. Our numerical results demonstrate gate calibration using Bayesian optimization based on state estimators (ROSE and PSE) with and without measurement noise of non-Markovian superconducting qubits, and we evaluate their
respective performance.
We also tested our method for different system parameter ranges, such as for energy, measurement strength, and correlation time for the non-Markovian environment. We find that our method makes qubit gate calibration faster with fewer measurements. 

Overall, we found that our calibration method is efficient, flexible, reliable, and easy to automate. Therefore, we expect it to become helpful for the operation of large-scale quantum computers. In future work, we plan to integrate the optimization and estimation of other parameters, such as qubit frequency, coupler frequency, and parameters related to the non-Markovian environment. We also plan to apply this method for multi-qubit calibration. Moreover, we expect the method to be useful not only for superconducting qubit systems but also for other platforms, such as ion trap qubits. Our work has the potential to pave the way for more efficient gate calibration procedures, as the size and complexity of quantum devices keep expanding.



\appendix


\section{Real-time optimal estimator}
The proposed real-time optimal state estimator (ROSE) of superconducting qubit has the form 
\begin{equation}
\begin{aligned}
d\hat{X}(t)=&A_{0}(t)dt+A_{2}(t)\hat{X}(t)dt\\&+K_{1}(\hat{X}(t),t)(dY(t)-d\hat{Y}(t))
\end{aligned}
\end{equation}
\begin{equation}
d\hat{Y}(t)=C\hat{X}(t)dt
\end{equation}
Above equation can also be written as:
\begin{equation}\label{Rose_X}
\begin{aligned}
d\hat{X}(t)=&A_{0}(t)dt+(A_{2}(t)-K_{1}(\hat{X}(t),t)C)\hat{X}(t)dt\\&+K_{1}(\hat{X}(t),t)dY(t)
\end{aligned}
\end{equation}
where we define an estimator of system's state $\hat{X}(t)=[\hat{x}(t),\hat{y}(t),\hat{z}(t)]^T$.

Imagine that if there is no noise we can directly know the two-level
state from $\hat{X}(t)$. Therefore, we need to minimize the error:
\begin{equation}
e_{1}(t)=X(t)-\hat{X}(t)
\end{equation}
The design purpose of the ROSE is to find the estimator gain $K_{1}(\hat{X}(t),t)$
that minimizes the estimation error covariance matrix $P_{1}(\hat{X}(t),t)=E[e_{1}e_{1}^{T}]$,
where E{[}\ensuremath{\centerdot}{]} is the expectation function.
So the rate of error is
\begin{equation}\label{A5}
\begin{aligned}
de_{1}(t)=&(A_{2}(t)-K_{1}(\hat{X}(t),t)C)e_{1}(t)dt+(G(X(t))\\&-\frac{K_{1}(\hat{X}(t),t)}{\sqrt{M}})dW(t)
\end{aligned}
\end{equation}
If the gain matrix $K_{1}(\hat{X}(t),t)$ is properly designed, then
the error will approach to zero with arbitrary decay. The fulfillment
of above requirement means that the system is to be asymptotically
stable. Now the rate of error covariance has the form
\begin{equation}\label{A6}
dP_{1}(\hat{X}(t),t)=E[de_{1}(t)e_{1}^{T}(t)+e_{1}(t)de_{1}^{T}(t)]
\end{equation}
Substituting equation (\ref{A5}) into equation (\ref{A6}), we have
\begin{eqnarray}\label{A7}
\begin{aligned}
\dot{P}_{1}(\hat{X}(t), t)=&(A_{2}(t)-K_{1}(\hat{X}(t), t)C)P_{1}(\hat{X}(t), t)\\&+P_{1}(\hat{X}(t), t)(A_{2}(t)-K_{1}(\hat{X}(t), t)C)^{T}\\&+(G(\hat{X}(t)-\frac{K_{1}(\hat{X}(t),t)}{\sqrt{M}})\centerdot\\&(G(\hat{X}(t)-\frac{K_{1}(\hat{X}(t),t)}{\sqrt{M}})^{T}
\end{aligned}
\end{eqnarray}
The objective is to minimize the error covariance, i.e.,$\min\limits_{K_1(\hat{X}(t), t)} \ tr \ \dot{P}{(\hat{X}(t), t)}$ w.r.t., estimator gain $K_1(\hat{X}(t), t)$, we have:
\begin{equation}
\begin{aligned}
\frac{\partial tr\dot{P}_{1}(\hat{X}(t), t)}{\partial K_{1}(\hat{X}(t), t)}&=-P_{1}(\hat{X}(t), t)C^{T}-P_{1}(\hat{X}(t), t)C^{T}\\&+2\frac{K_{1}(\hat{X}(t), t)}{M}-2\frac{G(\hat{X}(t))}{\sqrt{M}}=0
\end{aligned}
\end{equation}
Rearranging the above equation to find the estimator gain $K_{1}(\hat{X}(t), t)$,
we get
\begin{equation}\label{A9}
K_{1}(\hat{X}(t), t)C)=P_{1}(\hat{X}(t), t)C^{T}\sqrt{M}+G(\hat{X}(t))\sqrt{M}
\end{equation}
By using equations (\ref{A7}) and (\ref{A9}), the state dependent differential Riccati equation $\dot{P}_{1}(\hat{X}(t), t)$ can be expressed as
\begin{equation}\label{Rose_p}
\begin{aligned}
\dot{P}_{1}(\hat{X}(t), t)=&(A_{2}(t)-G(\hat{X}(t))C\sqrt{M})P_{1}(\hat{X}(t), t)\\&+P_{1}(A_{2}(t)-G(\hat{X}(t), t)C\sqrt{M})^{T}\\&-P_{1}(\hat{X}(t), t)C^{T}CP_{1}(\hat{X}(t), t)M
\end{aligned}
\end{equation}

\section{The second estimator without measurement noise}
The second estimator is designed to estimate the state of superconducting qubit without using the measurement noise. Considering this, first we take the derivative of Eq.~(\ref{qubitdy}) w.r.t., measurement strength $M$ and we have
\begin{eqnarray}
\begin{aligned}
d\frac{\partial X(t)}{\partial M}&=\frac{\partial A_2(t)}{\partial M}X(t)dt+A_2(t)\frac{\partial X(t)}{\partial M}dt\\&+\frac{\partial G(X(t))}{\partial M}dW(t)
\end{aligned}
\end{eqnarray}
we define $X_{P}(t)=\frac{\partial X(t)}{\partial M}$, and
\begin{equation}
dX_{P}=\frac{\partial A_2(t)}{\partial M}X(t)dt+A_2(t)X_{p}dt+\frac{\partial G(X(t))}{\partial M}dW(t)
\end{equation}
where ${X}_{P}(t)$ is the qubit state derivative induced by measurement. Now the second estimator of state $\hat{X}_{P}(t)$ of state ${X}_{P}(t)$ can be expressed as
\begin{eqnarray}\label{second_X}
\begin{aligned}
d\hat{X}_{P}(t)&=\frac{\partial A_2(t)}{\partial M}\hat{X}(t)dt+A_2(t)\hat{X}_{p}(t)dt\\
&+K_2(\hat{X}(t),\hat{X}_{P}(t),t)(dY-C\hat{X}(t)dt)
\end{aligned}
\end{eqnarray}
where $K_2(\hat{X}(t),\hat{X}_{P}(t),t)$ is the optimal gain matrix of second real-time state estimator. For simplicity we write $K_2(t)=K_2(\hat{X}(t),\hat{X}_{P}(t),t)$.

The error covariance matrix for the second estimator is $P_{2}(\hat{X}(t),\hat{X}_{P}(t),t)=E\left[e_2(t)e_{2}(t)^{T}\right]$, where $e_2(t)$ is the error, i.e., $e_2(t)=X_{p}(t)-\hat{X}_{p}(t)$. For simplicity we write $P_2(t)=P_2(\hat{X}(t),\hat{X}_{P}(t),t)$.
The derivation for getting the optimal gain $K_2(t)$ is similar to the real-time state estimator given in Appendix A and has the form as
\begin{equation}
K_{2}(t)=P_{2}(t)C^{T}\sqrt{M}+\sqrt{M}\frac{\partial G(X(t))}{\partial M}
\end{equation}
and by using this and the equation above we have
\begin{equation}
\begin{aligned}
\dot{P}_{2}(t)=&\left(A_{2}(t)-\frac{\partial G(\hat{X}(t))}{\partial M}\right)C\sqrt{M})P_{2}(t)\\&+P_{2}(t)\left(A_{2}(t)-\frac{\partial G(\hat{X}(t)}{\partial M}C\sqrt{M}\right)^{T}\\&-P_{2}(t)C^{T}CP_{2}(t)M\\&
+\frac{\partial A_2(t)}{\partial M}P_1(t)+P_{1}(t)^{T}\frac{\partial A^{T}_{2}(t)}{\partial M}
\end{aligned}
\end{equation}

One can see that the above ${P}_{2}(t)$ depends on the dynamics of $P_1(t)=P_{1}(\hat{X}(t),t)$ given in Eq. (\ref{Rose_p}). This illustrates that the ROSE (\ref{Rose_X}) and the second estimator (\ref{second_X}) must be executed simultaneously to get the estimation of $X_p(t)$.

\section{Bayesian Optimization}
The data Gaussian process $[g(\theta_1),...,g(\theta_j)]$ represents a function $g(\theta)$ to a set of proxy models $g(\theta)$. In order to limit the function value between $0$ and $1$ (to characterize fidelity), we take
\begin{equation}
    f(\theta)=\int_\infty^{g(\theta)} dy \frac{\exp(-y^2/2)}{\sqrt{2\phi}}
\end{equation}
At the same time we require that this family of functions to satisfy
\begin{equation}
\begin{aligned}
       k(\theta,\theta')&=\langle g(\theta)g(\theta')\rangle-\langle g(\theta)\rangle \langle g(\theta')\rangle\\&= V(1+\sqrt{5} |\theta-\theta'|/I + 5/3 )\exp(\theta,\theta')\exp(|\theta-\theta'|/I )
\end{aligned}
\end{equation}
This choice of kernel
ensures that any $f$ that occurs with nonvanishing probability is at least twice differentiable. The selection of $V$ and $I$ requires some data samples to determine over which the $f$ changes during optimization~\cite{shahriari}. Simply, it is obtained by minimizing $\text{log}(P(D|V,I))$, i.e., fitting out a Gaussian process that best fits out the existing observational data:
\begin{equation}\label{likehood}
    \text{log}(P(D|V,I))=-\frac{1}{2}D^Tk^{-1}D-\frac{1}{2}\text{log}|k|-c
\end{equation}
where $D$ is the experimental data $[f(\theta_0),...,f(\theta_i),...,f(\theta_m)]$. We vary the hyperparameters $V$ and $I$ of the kernel function $k$ to minimize the above Eqn.~(\ref{likehood}). Experimentally, we do not separate optimization parameters as in the previous protocol but optimize at the same time. We randomly select a few points in the control parameter space $(\theta_i)$, and then apply these control parameters to the pulse and measure the objective function, i.e., gate fidelity in Eq.~(\ref{gate fi}). Then we obtain the hyperparameters $V$ and $I$ by numerical optimization calculations.

After determining these two hyperparameters, we determine $P(f)$, so that we can predict the $f$-value of $f(\theta)$ for control parameters $\theta\rightarrow{(\alpha_{s},\phi)}$ that has not yet been measured. At the same time, we can theoretically deduce it
\begin{equation}
    P(f(\theta|\theta_i,D))=\mathcal{N} (\bar{f},\textrm{cov}(f))
\end{equation}
where $\mathcal{N}$ depicts the Gaussian distribution and
\begin{eqnarray}\label{upd1}
\begin{aligned}
\bar{f}=&k(\theta,\theta_i)k^{-1}(\theta_i,\theta_j)D\\
cov(f)=&k(\theta,\theta)-k(\theta,\theta_i)k^{-1}(\theta_i,\theta_j)k(\theta_i,\theta)
\end{aligned}
\end{eqnarray}
where $\bar{f}$ represents the average of $f(\theta)$ under the new parameters $\theta$ and $\textrm{cov}(f)$ is the variance.

Thus we can compute an interrogator function $\textrm{ac}(\theta)$ and has the form
\begin{eqnarray}\label{acfun}
\textrm{ac}(\theta) = \beta \ast \textrm{cov}(f)+f(\theta)
\end{eqnarray}
where $\beta$ determines the convergence speed and the local area size. 

Numerically, we look for the argument $\theta$ that makes the acquisition function $\textrm{ac}(\theta)$ larger. This process does not require experimental operations, only numerical optimization is needed. We take the search field segments, i.e., using the maximum method in each segment to find the largest $\textrm{ac}(\theta)$, and then we take a maximum of these $\textrm{ac}(\theta)$. This $\theta$ is the point we need to determine for our next experiment. We measured $f(\theta)$ to obtain new data points, so that the data set become $[f(\theta_0),...,f(\theta_i),...f(\theta_m)]$. Then we optimize the hyperparameters $V$ and $I$ again. Here we can choose not to optimize, or obtain a few new data points and then optimize. Although it depends on the needs of different control scenarios (in fact, even if not optimized, the results obtained are still relatively satisfactory). This way we get the new model $f$. At the same time, we can predict new $f(\theta)$ and calculate the new $\bar{f}$ and $\textrm{cov}(f)$ according to Eq. \eqref{upd1}. The new acquisition function $\textrm{ac}(\theta) =\beta\ast \textrm{cov}(f)+f(\theta)$ is then calculated.

\bibliographystyle{apsrev4-1}
\bibliography{References}

\end{document}